\documentclass[twocolumn,superscriptaddress,showpacs,prd,aps,amsmath,amssymb,nofootinbib]{revtex4-1}
\usepackage{graphicx}
\usepackage{bm}
\usepackage{color}
\usepackage{soul}
\usepackage{hyperref}
\usepackage{enumerate}
\usepackage{stackengine}

\newcommand{\beq}{\begin{equation}} 
\newcommand{\eeq}{\end{equation}} 
\newcommand{\beqn}{\begin{eqnarray}} 
\newcommand{\eeqn}{\end{eqnarray}}

\newcommand{\gmabu}{\gamma^{ab}}

\newcommand{\habu}{h^{ab}}
\newcommand{\habd}{h_{ab}}

\newcommand{\fabu}{f^{ab}}

\newcommand{\tD}{\tilde D}
\newcommand{\zD}{{\raise1.0ex\hbox{${}^{\ \circ}$}}\!\!\!\!\!D}
\newcommand{\alone}{{\raise0.5ex\hbox{${}^{\ 1}$}}\!\!\!\!\alpha}

\newcommand{\dl}{\delta}

\newcommand{\nalam}{\mathrel{\raise0.9ex\hbox{$^\lambda$}\mkern-14mu
\lower0.0ex\hbox{$\nabla$}}}

\newcommand{\tA}{\tilde A}

\newcommand{\zeroD}{{\raise1.0ex\hbox{${}^{\ \circ}$}}\!\!\!\!\!D}

\newcommand{\zLap}{{\raise1.0ex\hbox{${}^{\ \circ}$}}\!\!\!\!\Delta}
\newcommand{\zna}{{\raise1.0ex\hbox{${}^{\ \circ}$}}\!\!\!\!\!\nabla}
\newcommand{\zS}{{\raise1.0ex\hbox{${}^{\ \circ}$}}\!\!\!\!\!S}

\newcommand{\cocal}{\textsc{cocal}}

\newcommand{\illinois}{\textsc{Illinois GRMHD}}

\newcommand{\GA}{\alpha}
\newcommand{\GB}{\beta}
\newcommand{\GG}{\gamma}
\newcommand{\GD}{\delta}
\newcommand{\GE}{\epsilon}

\newcommand{\GR}{\rho}
\newcommand{\GS}{\sigma}

\newcommand{\GC}{\psi}

\newcommand{\GO}{\omega}

\newcommand{\GP}{\phi}

\newcommand{\GH}{\eta}
\newcommand{\GU}{\theta}
\newcommand{\nl}{\rm \scriptscriptstyle NL}
\newcommand{\tf}{\rm \scriptscriptstyle TF}
\newcommand{\ks}{\rm \scriptscriptstyle KS}
\newcommand{\pd}{\partial}
\newcommand{\be}{\begin{equation}}
\newcommand{\ee}{\end{equation}}

\newcommand{\Bn}{\mathbf{n}}
\newcommand{\Bt}{\mathbf{t}}

\newcommand{\Bph}{\boldsymbol{\GP}}

\newcommand{\TD}[2]{\tilde{#1}_{#2}}
\newcommand{\TU}[2]{\tilde{#1}^{#2}}
 
\newcommand{\TDD}[3]{\tilde{#1}_{#2 #3}} 
\newcommand{\TDU}[3]{\tilde{#1}_{#2}^{\ #3}} 
\newcommand{\TUD}[3]{\tilde{#1}^{#2}_{\ #3}} 
\newcommand{\TWD}[2]{( \tilde{\mathbb L} \tilde{W} )_{#1 #2}} 
 
\newcommand{\TBD}[2]{( \tilde{\mathbb L} \tilde{\GB} )_{#1 #2}}

\newcommand{\tW}{\tilde{W}}
\newcommand{\tGS}{\tilde{\GS}}
\newcommand{\tGG}{\tilde{\GG}}
\newcommand{\tGB}{\tilde{\GB}}

\newcommand{\tRs}{\tilde{\mathcal{R}}}

\newcommand{\C}[3]{C^{#1}_{\ {#2}{#3}}}

\newcommand{\flap}{{\raise0.8ex\hbox{${}^{\ \scalebox{0.7}{$\circ$}}$}}\hspace{-6.3pt}\Delta}

\newcommand{\fD}{{\raise0.8ex\hbox{${}^{\ \scalebox{0.7}{$\circ$}}$}}\hspace{-6.6pt}D}

\begin{document}
\title{Complete initial value spacetimes containing black holes in general relativity:
Application to black hole-disk systems}
\author{Antonios Tsokaros}
\affiliation{Department of Physics, University of Illinois at
  Urbana-Champaign, Urbana, IL 61801}    
\author{K\=oji Ury\=u}
\affiliation{Department of Physics, University of the Ryukyus, Senbaru, Nishihara, Okinawa 903-0213, Japan}
\author{Stuart L. Shapiro}
\affiliation{Department of Physics, University of Illinois at
  Urbana-Champaign, Urbana, IL 61801}
\affiliation{Department of Astronomy \& NCSA, University of
  Illinois at Urbana-Champaign, Urbana, IL 61801}

%%%%%%%%%%%%%%%%%%%%
%%%   ABSTRACT   %%%
%%%%%%%%%%%%%%%%%%%%
\begin{abstract}
We present a new initial data formulation to solve the \textit{full set} of Einstein equations for 
spacetimes that contain a black hole under general conditions.
The method can be used to construct complete initial data for spacetimes (the full metric)
that contain a black hole. Contrary to most current studies the formulation requires minimal assumptions.
For example, rather than imposing the form of the spatial conformal metric we impose 3 gauge conditions
adapted to the coordinates describing the system under consideration.  %physical model we are constructing. 
For stationary, axisymmetric spacetimes our method
yields Kerr-Schild black holes in vacuum and rotating equilibrium neutron stars. 
We demonstrate the power of our new method by solving for the first time the whole system 
of Einstein equations for a nonaxisymmetric, self-gravitating torus in the presence of a 
black hole. The black hole has
dimensionless spin $J_{\rm bh}/M_{\rm bh}^2=0.9918$, a rotation axis  tilted at 
a $30^\circ$ angle with respect to the angular momentum of the disk, and a mass of 
$\sim 1/5$ of the disk. 
\end{abstract}

\pacs{04.25.D-, 04.25.dg, 47.75.+f} % PRD pacs

\maketitle

%%%%%%%%%%%%%%%%%%%%%%
%%%  Introduction  %%%
%%%%%%%%%%%%%%%%%%%%%%
\section{Introduction}

Although the term ``black hole'' was coined fairly recently by John Wheeler 
52 years ago, and its physical significance was questioned earlier by Einstein
himself, it turns out that $21$st century physics will be dominated
by these extraordinary objects. The spectacular first detection by the Laser 
Interferometer  Gravitational-Wave Observatory (LIGO) of a merging binary 
black hole system \cite{1stBBH16}, as well as the nine follow-up detections,
unequivocally confirmed their existence and many of their properties. 
%Black holes are expected to arise 
%in binary neutron star mergers like \cite{1stBNS17} as well as in the center 
%of galaxies. 

Accurate modeling of black hole spacetimes requires initial data that 
precisely describe the systems under consideration. Over the years three main 
different ideas have been heavily employed to address this problem. These are 
the conformal transverse  traceless (CTT) decomposition \cite{York71,York72,Pfeiffer03}, 
the puncture method \cite{Brandt97}, and the conformal thin-sandwich (CTS) approach 
\cite{York99} (see \cite{BS10} for summary and discussions). 
Many variants of these formulations exist but of particular importance
is the so-called Isenberg-Wilson-Mathews \cite{IWM} formulation, whose strength stems 
from its simplicity and versatility, as it is used for black hole as well as neutron star 
spacetimes.

All methods described above solve a subset of the Einstein equations. One common 
characteristic is that they assume the form of the spatial conformal metric which is associated 
with the true dynamical degrees of freedom of the gravitational field \cite{York72}.
In \cite{Vasset09} the authors used a constrained scheme presented in \cite{Bonazzola04}
for the Einstein equations  to solve for the conformal metric as well. In doing so they discovered 
solutions that better satisfy the Einstein equations (at least in comparison to conformally 
flat solutions) and reach dimensionless spins up to $J_{\rm bh}/M_{\rm bh}^2\sim 0.85$. Their solutions 
displayed Kerr-like properties but they were not expressed in any of the well-known coordinates, like the
Kerr-Schild ones \cite{Kerr63}, which are known to yield high-spin initial data and exhibit good behaviour in 
evolution simulations.

In this paper we present a new formulation and a new code within the \cocal{}
(Compact Object CALculator) project \cite{Uryu12} that solves \textit{all}
the Einstein equations in a self-consistent manner and achieves the following:
(1) In the absence of matter our code can reproduce the exact Kerr-Schild solution, even for high
spins. No assumptions on axisymmetry are imposed and therefore this is the first \textit{generic}
 3-d method that obtains an exact Kerr solution and can be applied with minimal changes to a broad range of 
\textit{nonaxisymmetric} problems, such as tilted disks or binary systems.
(2) The domain of the solution extends inside the apparent horizon, which is well-suited for
evolution simulations. (3) In the presence of massless disks around the black hole our code 
reproduces well-known solutions (e.g. \cite{Chakrabarti85,Villiers03}). 
(4) The first self-consistent, tilted black hole-torus solutions
are presented %that satisfy not only the constraint equations but the \textit{total} Einstein system.
that solve for the \textit{total} spacetime metric.
In addition, these are the highest mass ratio and black hole spin solutions
constructed for black hole-torus systems to date.
We present a solution with a 
spinning black hole whose dimensionless spin is $J_{\rm bh}/M_{\rm bh}^2=0.9918$, has
an angle with respect to the angular momentum of the torus of $\GU=30^\circ$, while the torus has rest mass 
approximately five times the black hole mass. 

Tilted disk-black hole systems can be produced in the merger 
of black hole-neutron star systems where the spin of the black hole is tilted with respect to the total angular 
momentum of the system \cite{Foucart13,Kawaguchi15}. 
Tilted black holes may also arise in massive disks in active galactic nuclei and quasars \cite{Natarajan99}.

In the following, greek letters denote spacetime indices while latin letters indicate spatial ones. 
We adopt units with $G=c=M_\odot=1$, unless otherwise stated.

%%%%%%%%%%%%%%%%%%%%%
%%%  Formulation  %%%
%%%%%%%%%%%%%%%%%%%%%
\section{Formulation for gravity}

We use the standard $3+1$ formalism to express spacetime $\mathcal{M}=\mathbb{R}\times\Sigma_t$  
as a foliation of three dimensional spacelike hypersurfaces $\Sigma_t$ ($t$ labels the hypersurface)
with spatial coordinates ${x^i}$ and unit normal vector $n^\mu$.
Points with the same values of $x^i$ in neighboring hypersurfaces
are connected with a timelike vector $t^\mu$ that can be decomposed as $t^\mu:=\GA n^\mu + \GB^\mu$,
where $\GA$ is the lapse and $\GB^\mu$ is the (spatial) shift vector. The first fundamental form of the
hypersurfaces is $\GG_{\GA\GB}:=g_{\GA\GB}+n_\GA n_\GB$, and the full spacetime
line element
is $ds^2=-\GA^2 dt^2 + \GG_{ij}(dx^i+\GB^i dt)(dx^j+\GB^j dt)$. A conformal geometry is introduced by
setting $\GG_{ij}:=\GC^4 \TDD{\GG}{i}{j}$. We further define $\TDD{\GG}{i}{j}:=f_{ij}+h_{ij}$, where 
$f_{ij}$ is the flat metric in arbitrary coordinates and $h_{ij}$ the nonflat contributions 
\textit{which we wish to evaluate} together with the rest of the potentials, $\GC,\GA,\GB^i$, 
that are computed traditionally.
For the conformal geometry we assume ${\rm det}(\TDD{\GG}{i}{j})=\tilde{\GG}={\rm det}(f_{ij})$.

The initial data are the 3-metric $\GG_{ij}$ and the extrinsic curvature
$K_{ij}=-\frac{1}{2}\mathcal{L}_{\Bn}\GG_{ij}$ 
($\mathcal{L}_{\Bn}$ denotes the Lie derivative with respect to the unit normal 
$n^\GA$) which is further decomposed as
$K_{ij}=A_{ij}+\frac{1}{3}\GG_{ij}K$, where $K$ is its trace and $A_{ij}$ its
tracefree part. The conformal tracefree part of the extrinsic curvature is
defined as $\tA_{ij}=\GC^{-4}A_{ij}$ and we introduce the decomposition 
\be
\TDD{A}{i}{j} = \tilde{A}^{\ks}_{ij} + \tGS\TWD{i}{j}  ,  \label{eq:ctt}
\ee
where $\tilde{A}^{\ks}_{ij}$ is the Kerr-Schild part, $\tW_i$ an unknown spatial 
vector, and $\tGS$ a scalar. $\tilde{\mathbb L}$ is the conformal Killing 
operator:
$\TWD{i}{j}=\TD{D}{i}\TD{W}{j}+\TD{D}{j}\TD{W}{i}-\frac{2}{3}\TDD{\GG}{i}{j}\TD{D}{k}\TU{W}{k}$.

A Kerr black hole spacetime in Kerr-Schild coordinates \cite{Kerr63} can be written as
$ds^2=(\GH_{\GA\GB}+2 H l_\GA l_\GB)dx^\GA dx^\GB$, where $H=mr^3/(r^4+(a_i x^i)^2)$,
$r^2=(x^ix_i -a^ia_i)/2+\sqrt{(x^ix_i-a^ia_i)/4+(a^ix_i)^2}$, and                
$l_\GA=(1,l_i)$ with $l_i=x^j(r^2 \GD_{ij}+r\GE_{ijk}a^k +a_i a_j)/(r(r^2+a^2))$.
Note $r^2\neq x_ix^i = \hat{r}^2$. Here $x_i=x^i$ and $a^i$                      
is the spin of the black hole, $a^2=a_ia^i$, and $l_\GA$ is a null vector, both with respect to the spacetime
metric $g_{\GA\GB}$ as well as the Minkowski metric $\GH_{\GA\GB}$. The $3+1$ quantities of an
arbitrarily spinning black hole in Kerr-Schild coordinates are                   
$\GC_{\ks}=(1+2H)^{1/12}$, $\GA_{\ks}=1/\sqrt{1+2H}$,                            
$\GB^i_{\ks}=2H\GA^2 l^i$, $\GG_{ij}^{\ks}=\GD_{ij}+2 H l_i l_j$ and therefore   
$h_{ij}^{\ks}=\GC^{-4}_{\ks}(\GD_{ij}+2H l_i l_j)-\GD_{ij}$.                     
Using the 3+1 quantities above one can compute $\tilde{A}^{\ks}_{ij}$ which
appears in Eq. (\ref{eq:ctt}). The trace of the extrinsic curvature in 
Kerr-Schild coordinates is               
\be                                                                              
K_{\ks}=\frac{2H\GA^3_{\ks}}{r}\left(1+H+\frac{2H^2 r}{m}\right).   \label{eq:trk}
\ee  
and in our calculations we assume $K=K_{\ks}$.

Taking combinations of the projections of the Einstein equations 
($(G_{\mu\nu}-8\pi T_{\mu\nu})n^\mu n^\nu=0$, $(G_{\mu\nu}-8\pi T_{\mu\nu})n^\mu \GG^{\nu}_{\ i}=0$,
$(G_{\mu\nu}-8\pi T_{\mu\nu})\GG^{\mu}_{\ i} \GG^{\nu}_{\ j}=0$) onto the spatial
hypersurface \cite{Shibata04,Uryu09} one can arrive at a set of elliptic equations 

\begin{widetext}
\begin{eqnarray}
\flap \GC & = & -h^{ij}\fD_i\fD_j\GC + \tGG^{ij}\C{k}{i}{j}\fD_k\GC +\frac{1}{8}\GC\tRs - 
\frac{\GC^5}{8}\left(\tA_{ij}\tA^{ij}-\frac{2}{3}K^2\right) - 2\pi \GR_{\rm H} \GC^5 ,  \label{eq:HC}\\
 \flap \tW_i & = & -h^{ab}\fD_a\fD_b\tW_i + 
                    \tGG^{ab} [\fD_a(\C{m}{b}{i}\tW_m) + \C{m}{a}{b}\TD{D}{m}\tW_i + \C{m}{a}{i}\TD{D}{b}\tW_m]
                    -\frac{1}{3}\fD_i \fD_j \tW^j  - \tRs_{ij}\tW^j  \nonumber \\
             & - &   \frac{1}{\tGS}\tilde{D}^j\tA_{ij}^{\ks} - \frac{1}{\tGS}\tilde{D}^j\tGS\TWD{i}{j} 
                    - \frac{6}{\GC\tGS}\tilde{D}^j\GC\tA_{ij}  
                    + \frac{2}{3\tGS}\fD_i K + \frac{8\pi}{\tGS} j_i , \label{eq:wvec}  \\
\flap \tGB_i & = & -h^{ab}\fD_a\fD_b\tGB_i + 
                    \tGG^{ab} [\fD_a(\C{m}{b}{i}\tGB_m) + \C{m}{a}{b}\TD{D}{m}\tGB_i + \C{m}{a}{i}\TD{D}{b}\tGB_m]
                    -\frac{1}{3}\fD_i\fD_j \tGB^j - \tRs_{ij}\tGB^j 
                    -\frac{2\GA^2}{\GC^6}\TDU{A}{i}{j}\fD_j\left(\frac{\GC^6}{\GA}\right)  \nonumber \\ 
             & + &   \frac{4\GA}{3}\fD_i K + \tGG^{jm}\TD{D}{j} \tilde{u}_{im} + 16\pi\GA j_i \label{eq:MC} ,   \\
\flap (\GA\GC) & = & -h^{ij}\fD_i\fD_j(\GA\GC) + \tGG^{ij}\C{k}{i}{j}\fD_k(\GA\GC) + \frac{\GA\GC}{8}\tRs
                     +\GA\GC^5\left(\frac{7}{8}\tA_{ij}\tA^{ij}+\frac{5}{12}K^2\right)
                      -\GC^5\mathcal{L}_{\GA \Bn}K + 2\pi\GA\GC^5(\GR_{\rm H}+2S) ,  \label{eq:trdKijdt} \\
%\flap h_{ij} & = & -\frac{1}{3}\tGG_{ij}\fD_k h_{ab}\fDu{k}h^{ab} + \frac{2}{3}\tGG_{ij}\fDu{k}\C{a}{a}{k} 
\flap h_{ij} & = & -\frac{1}{3}\tGG_{ij}\fD_k h_{ab}\fD^k h^{ab} + \frac{2}{3}\tGG_{ij}\fD^k \C{a}{a}{k} 
                   +2\left[\tRs_{ij}^{\ks} + \tRs_{ij}^{\nl} - 8\pi S_{ij} 
                      + \GC^4\left(\frac{1}{3}K\tA_{ij}-2\tA_{ik}\TUD{A}{k}{j}\right)\right.  \nonumber \\
             & + &  \left.\frac{1}{\GA\GC^2}\left(-\fD_i\fD_j(\GA\GC^2) + \C{k}{i}{j}\fD_k(\GA\GC^2)
                          +4\fD_i(\GA\GC)\fD_j\GC + 4\fD_i\GC\fD_j(\GA\GC)\right) 
                          -\frac{1}{\GA} \mathcal{L}_{\GA\Bn}(\GC^4\tA_{ij})    \right]^{\tf} ,  \label{eq:tfdKijdt}
\end{eqnarray}
\end{widetext}
for the eleven metric potentials $\GC,\GA,\tGB_i, h_{ij}$ and the three
auxiliary components of $\tW_i$.
We define $h^{ij}$ through $\tGG^{ij}=f^{ij}+h^{ij}$
where $\tGG^{ij},f^{ij}$ the inverses of $\tGG_{ij},f_{ij}$.
The covariant derivatives associated with $\GG_{ij},\tGG_{ij},f_{ij}$ are respectively $D, \tilde{D}$, and $\fD$. 
The symbol $\flap$ means $\flap=\fD_k \fD^{k}$.
It is $D_i\GB^k =\tD_i\GB^k + \tilde{C}^k_{\ ij}\GB^j$ and $\tD_i \GB^k=\fD_i\GB^k + C^k_{\ ij}\GB^j$
where
$\tilde{C}^k_{\ ij}=\frac{2}{\GC}(\tGG^k_{\ i}\tD_j\GC + \tGG^k_{\ j}\tD_i\GC - \tGG_{ij}\tGG^{km}\tD_m\GC)$
and 
$C^k_{\ ij} = \frac{1}{2}\tGG^{km}(\fD_i h_{mj}+\fD_j h_{mi}-\fD_m h_{ij})$.      
Contraction on the first two indices results to $C^k_{\ kj}=\frac{1}{2\tGG}\fD_j\tGG$ and
$\tilde{C}^k_{\ kj}=\frac{1}{2\GG}\tD_j\GG$. For $\tGG=1$, as in our computations, $C^k_{\ kj}=0$.
In the case where Cartesian coordinates are used for the flat metric, $f_{ij}=\GD_{ij}$ then
$\fD$ is the usual partial derivative $\pd$, and $\flap$ the Laplacian in Cartesian coordinates.
The superscript $^{\tf}$ means the trace-free part.
The conformal shift is defined as $\TU{\GB}{i}=\GB^i$ and thus $\TD{\GB}{i}=\TDD{\GG}{i}{j}\TU{\GB}{j}=\psi^{-4} \GB_i$.
The matter sources that appear on the right-hand side of Eqs. 
(\ref{eq:HC})-(\ref{eq:tfdKijdt}) are  $\GR_{\rm H}=T_{\mu\nu}n^\mu n^\nu$,
$j_i =-T_{\mu\nu}\GG^\mu_{\ i} n^\nu$, $S=T_{\mu\nu}\GG^{\mu\nu}$, and 
$S_{ij}=T_{\mu\nu}\GG^\mu_{\ i}\GG^\nu_{\ j}$.

Eq. (\ref{eq:HC}) is the Hamiltonian constraint, Eq. (\ref{eq:wvec}) as well as 
Eq. (\ref{eq:MC}) are the momentum constraints, Eq. (\ref{eq:trdKijdt}) is the 
spatial trace of the Einstein's equation for $\pd_t K_{ij}$ combined with the 
Hamiltonian constraint, and Eq. (\ref{eq:tfdKijdt}) is the spatial tracefree 
part of Einstein's equation. 
Equations (\ref{eq:wvec}), (\ref{eq:MC}) imply that we solve for the momentum 
constraint twice. This idea has been used successfully in \cite{SU06} and there 
are two reasons for adopting this method. First by introducing Eq. 
(\ref{eq:ctt}) one has now two expressions for the conformal traceless extrinsic
curvature, the second being
\be
\tA_{ij} = \frac{1}{2\GA}[ \TBD{i}{j} - \tilde{u}_{ij} ],\qquad  \tilde{u}_{ij}=[\pd_t \tGG_{ij}]^{\tf},
\label{eq:taij}
\ee
which involves the shift vector. Solving for $\GB^i$ is necessary since it will
be used in the computation of $\mathcal{L}_{\GA \Bn}K$ in Eq. 
(\ref{eq:trdKijdt}) and $\mathcal{L}_{\GA\Bn}(\GC^4\tA_{ij})$ in Eq. 
(\ref{eq:tfdKijdt}). The second reason is that the introduction of Eq.
(\ref{eq:ctt}) (i.e. resolving the momentum constrain for $\tW_i$) enables us
to obtain \textit{apparent horizon penetrating solutions}. 
In particular since in our
method we use excision, the use of this extra decomposition makes possible the
use of grids that excise a region \textit{inside the apparent horizon}, which
facilitates the evolution of our systems. Without decomposition (\ref{eq:ctt}) 
the system of Eqs. (\ref{eq:HC}), (\ref{eq:MC}), (\ref{eq:trdKijdt}), 
(\ref{eq:tfdKijdt}) with Kerr-Schild inner boundary conditions and extrinsic 
curvature given by Eq. (\ref{eq:taij}) converges only when the excised region 
is \textit{outside} the apparent horizon. The faster the black hole spins the 
further out one has to perform the excision. 

In this work we choose $\pd_t\tGG_{ij}=0$ and therefore $\tilde{u}_{ij}=0$.
Similarly we assume $\pd_t\tA_{ij}=\pd_t K=0$.
This is consistent with stationary systems like rotating stars or a Kerr black 
hole. In binary systems where one typically assumes a helical symmetry, 
$k^\GA=t^\GA + \Omega\GP^\GA$, a better choice would be 
$\mathcal{L}_{\mathbf{k}}\tGG_{ij}=0=\mathcal{L}_{\mathbf{k}}\tA_{ij}$ which results to 
$\tilde{u}_{ij}=-\Omega (\tilde{\mathbb L}\tilde{\GP})_{ij}$.

Another important term in our system is the one that involves $\tRs_{ij}$, the 3-d
Ricci tensor associated with the conformal geometry $\tGG_{ij}$. One can show \cite{Shibata04} that
\be
\tRs_{ij} = -\frac{1}{2} \flap h_{ij} + \tRs_{ij}^{\ks} + \tRs_{ij}^{\nl} , \label{eq:tRij}
\ee
where 
\begin{eqnarray}
\tRs_{ij}^{\ks} & = & -\frac{1}{2}(f_{ik}\fD_j F^k + f_{jk}\fD_i F^k),  \label{eq:tRks}  \\
\tRs_{ij}^{\nl} & = & -\frac{1}{2}(h^{ab}\fD_a\fD_b h_{ij} + \fD_i h^{ab}\fD_b h_{aj} + \fD_j h^{ab} \fD_a h_{ib}) \nonumber \\ 
                & - &  \frac{1}{2}[\fD_i(h_{kj}F^k) + \fD_j(h_{ik}F^k)]  \nonumber  \\
                & - & \fD_i C^k_{\ kj} + C^k_{\ km}C^m_{\ ij} + F^k C_{kij} - C^k_{\ im}C^m_{\ kj} ,  \label{eq:rijnl}
\end{eqnarray}
and $F^i=\fD_a\tGG^{ia}$.
Notice that the terms $\tRs_{ij}^{\ks}$,  $\tRs_{ij}^{\nl}$ also enter into Eq. (\ref{eq:tfdKijdt}), which we
discuss below. The nonlinear term $\tRs_{ij}^{\nl}$ is second order in $h_{ij}$
and therefore smaller than
the first order terms $\tRs_{ij}^{\ks}$ and $\flap h_{ij}$ in Eq.
(\ref{eq:tRij}). The term $\tRs_{ij}^{\ks}$
involves the gauge functions $F^i$ which are identical to the $\tilde{\Gamma}^i$ in the 
Baumgarte-Shapiro-Shibata-Nakamura (BSSN) formulation \cite{BS10}. For initial
data, in order for the 
whole system (\ref{eq:HC}-\ref{eq:tfdKijdt}) to converge, \textit{these functions must be fixed} \cite{Shibata04}. 
For rotating stars
the Dirac gauge condition $F^i=0$ was used \cite{Lin2006,Uryu16}, which in the case of stationary and axisymmetric 
problems is also known to yield solutions numerically identical to the exact ones \cite{Lin2006}.
For binary neutron star systems the same Dirac gauge condition $F^i=0$ was used in \cite{Uryu09,Uryu06} to produce
the most accurate initial data, especially for the late inspiral binaries. Similarly \cite{Vasset09} applied
that gauge for single black hole spacetimes.
Here, since we want to be able to retrieve the Kerr-Schild black hole, we set 
\be
F^j=\fD_i h^{ij}_{\ks} .   \label{eq:fi}
\ee
where $h^{ij}_{\ks}$ are the exact Kerr-Schild potentials.
Equation (\ref{eq:fi}) are gauge conditions that are related to our freedom in choosing spatial coordinates. 
Setting $F^i=0$ for black holes may yield solutions qualitatively close to Kerr-Schild, but spins only 
up to $0.85$ \cite{Vasset09}.

Imposing conditions (\ref{eq:fi}) to the solutions of the system Eq. (\ref{eq:HC})-(\ref{eq:tfdKijdt}),
and thereby having a self-consistent iteration scheme, an adjustment is necessary for the $h_{ij}$.
Following \cite{Uryu09}, [or \cite{Uryu16} Eq. (29-32)] gauge vector potentials $\xi^a$ introduced in the 
transformation 
\be
\dl \gmabu \rightarrow \dl \gmabu \,-\, \fD^a\xi^b \,-\, \fD^b\xi^a, 
\label{eq:gauge_transf}
\ee
are used to adjust $\habu$ as 
\beq
\habu{}' =  \habu \,-\, \fD^a\xi^b \,-\, \fD^b\xi^a \,+\, \frac23\fabu \fD_c\xi^c, 
\label{eq:gauge_hab}
\eeq
where now $\habu{}'$ are chosen to satisfy the condition $\fD_b \habu{}'=F^a$ given by (\ref{eq:fi}).  
The gauge vector potentials $\xi^a$ are solved from the elliptic equations, 
\be
 \flap \xi^a \,=\, \fD_b \habu - \frac13 \fD^a \fD_b\xi^b - F^a ,
\label{eq:gauge}
\ee
and then $\habd$ are replaced by Eq.~(\ref{eq:gauge_hab}).  

In our method we use excision \cite{Tsokaros07} with inner boundary conditions
being the exact Kerr-Schild values on some excised sphere, chosen inside the 
outer black hole horizon. For outer boundary conditions we use ones that lead 
to an asymptotically flat spacetime. The augmented system of the 17 elliptic 
equations (\ref{eq:HC})-(\ref{eq:tfdKijdt}), (\ref{eq:gauge}) with 
$\tilde{\GS}=1/(2\GA)$ and zero boundary conditions
for the gauge potentials and the vector $\TD{W}{i}$ converges smoothly in 
vacuum or in the presence of matter (like a massive disk), even for near 
maximally-spinning black holes. In a typical iteration
we first solve Eq. (\ref{eq:wvec}) to obtain $\tW_i$, then $\TDD{A}{i}{j}$ is 
constructed  through Eq. (\ref{eq:ctt}) which is then used in the right-hand 
side of Eqs. (\ref{eq:HC})-(\ref{eq:tfdKijdt}) to compute the rest of the 
potentials. 
We want to emphasize that any solution of our method not only satisfies the 
constraint equations but it \textit{also solves for the conformal geometry, thus 
providing a way to control the gravitational wave content of the initial data 
in a self-consistent way}.

%%%%%%%%%%%%%%%%%%%%%
%%%     Disk      %%%
%%%%%%%%%%%%%%%%%%%%%
\section{Formulation for the fluid}

As a first application of our new formulation we compute massive disks in the presence of 
tilted black holes. Such systems will inevitably have rich behavior as they \textit{cannot} be in 
equilibrium \cite{Rezzolla10,Foucart11,Mewes16}. 

We assume that the stress energy tensor is described by a perfect fluid with 4-velocity $u^\GA$:
$T^{\GA\GB}=\GR h u^\GA u^\GB + pg^{\GA\GB}$, where $h$ is the specific enthalpy, $\GR$ the rest-mass
density and $p$ the fluid pressure. It is $\GR h=\GR+p$. Bianchi identity together with the $1^{st}$ law
of thermodynamics $\GR dh=\GR Tds+dp$, implies 
$\nabla_\GA T^{\GA\GB}=\GR u^\GB \GO_{\GA\GB}+hu_\GA\nabla_\GB(\GR u^\GB)-\GR T\nabla_\GA s =0$. Here
$\GO_{\GA\GB}=\nabla_\GA(hu_\GB) - \nabla_\GB(hu_\GA)$ is the relativistic vorticity. By assuming conservation
of rest-mass $\nabla_\GA(\GR u^\GA)=0$ and an isentropic flow one arrives at the relativistic Euler equation 
$u^\GB \GO_{\GA\GB}=0$ \cite{Uryu16}. 

The approximate symmetries that will be invoked will determine the fluid motion. One 
approximation that can be adopted is to extend the quasistationarity condition 
of the gravitational fields to the initial \textit{fluid variables} as well. Thus assume 
$\mathcal{L}_{\Bt}(h u_\GA)=\mathcal{L}_{\Bt} \GR =0$. 
We can also assume that the fluid \textit{motion} is axisymmetric about the fluid axis, which 
we take to be the z-axis.
Thus $\mathcal{L}_{\Bph}(hu^\GA)=\mathcal{L}_{\Bph}\GR=0$, where $\GP^i$ the generator of rotations 
around the $z$-axis. Under these assumptions, and the fact that 
the matter 4-velocity can be written as $u^\GA=u^tk^\GA$ with $k^\GA=t^\GA+\Omega\GP^\GA$, the
Euler equation becomes $\nabla_\GA\ln\frac{h}{u^t}+u^t u_\GP\nabla\Omega=0$.

Assuming that the combination  $u^t u_\GP$ is a function of the angular velocity $\Omega$ one 
arrives at
\be
\frac{h}{u^t} e^{\int j(\Omega) d\Omega} \ = \ \mathcal{E},\qquad  j(\Omega)=u^t u_\GP , 
\label{eq:euler}
\ee
where $\mathcal{E}$ a constant. In terms of the gravitational variables it is
$j(\Omega)=\GG_{ij}\GO^i\GP^j/(\GA^2-\GG_{ij}\GO^i\GO^j)$, with $\GO^i=\GB^i+\Omega\GP^i$. 

For this work we assumed a barotropic fluid EoS with $p=K\GR^\Gamma$, where $K,\Gamma$ are constants, 
but our code  can compute more exotic EoSs as in \cite{Zhou12} or even in a tabulated form. 
Given an EoS, as well as a differential rotation form for $j(\Omega)$ \cite{Uryu17}, the algebraic equation 
(\ref{eq:euler}) must be solved together with (\ref{eq:HC})-(\ref{eq:tfdKijdt}),
and (\ref{eq:gauge}) in order to compute the fluid variables $u^t,\GR$ (alternatively $h$ or $p$) and
the constant $\mathcal{E}$ together with the gravitational variables.

%%%%%%%%%%%%%%%%%%%%%%%%%%
%%%  Numerical method  %%%
%%%%%%%%%%%%%%%%%%%%%%%%%%
\section{Numerical implementation}

For the numerical solution of the Poisson-type of equations, 
Eqs. (\ref{eq:HC})-(\ref{eq:tfdKijdt}), and (\ref{eq:gauge}), 
we use the Komatsu-Eriguchi-Hachisu (KEH)
method for black holes, which was first developed in \cite{Tsokaros07} and implemented
within the {\cocal} code in \cite{Uryu12}. The Green's functions used in the representation
formula match the boundary conditions that we impose on our variables,
$\{ \GC, \TD{\GB}{i}, \GA, h_{ij}, \xi^i,\TD{W}{i} \}$, and in the present calculation are the
Dirichlet-Dirichlet functions (for all variables), Eq. (B8) in \cite{Tsokaros07}. 
A single spherical $(\hat{r},\hat{\GU},\hat{\GP})$ grid is used, 
identical to the black-hole grids of \cite{Tsokaros07}, with uniform intervals in $\hat{\GU},\hat{\GP}$
and non-uniform intervals in $\hat{r}$. In the solutions presented
here we used $N_r \times N_\GU \times N_\GP = 660\times 48\times 48$ intervals that cover the
whole space $\hat{r}\in[\hat{r}_a,\hat{r}_b], \hat{\GU}\in[0,\pi]$, and $\hat{\GP}\in[0,2\pi]$. 
Here $\hat{r}_a$ denotes the excised sphere inside the horizon and $\hat{r}_b= 10^5 m$.
Convergence studies in the new formulation will be presented elsewhere \cite{atsok18}.

Apart from the isolated Kerr solution we have computed as a check massless axisymmetric disks in the 
presence of a Kerr black hole using the differential law $\Omega = k \ell^\GA $ \cite{Chakrabarti85} 
where $k,\GA$ constants and $\ell=-u_\phi/u_t$ the specific angular momentum.
Note that $j(\Omega)=u_t u^\GP = \ell/(1-\Omega \ell)$.
For a black hole spin $a/m=0.9$, differential law parameter $\GA=-17/3$, polytropic index $\Gamma=1.\bar{4}$,
and inner point disk characteristics $\ell_{\rm in}/m=3.313$, $r_{\rm in}/m=6$
our solution shows excellent agreement (maximum density agrees to four significant digits) 
with a calculation of \cite{Farris11}, used to generate an equilibrium solution prescribed in \cite{Villiers03}
and constructed via the \illinois{} code \cite{Etienne10}.

%%%%%%%%%%%%%%%%%%%%%%%%%%%%%%%%%%%%%%%%
%%%  Tilted black hole-torus system  %%%
%%%%%%%%%%%%%%%%%%%%%%%%%%%%%%%%%%%%%%%%
\section{Tilted black-hole-torus system}
\label{sec:bht}

The first self-gravitating black hole-toroidal systems have been computed by Nishida 
and Eriguchi, \cite{Nishida94}, (see also Stergioulas \cite{Stergioulas11}),
while more recently, using 
different methods, by Ansorg and Petroff \cite{Ansorg05}, as well as Shibata
\cite{Shibata07} (see also Karkowski et al. \cite{Karkowski18}). All authors 
computed equilibria by solving the 2-d problem of stationary and axisymmetric Einstein equations. 

With our new method we computed sequences of full 3-d nonaxisymmetric solutions of self-gravitating tori
around tilted black holes. In order to do that we fix the inner point of the torus along the x-axis
(here we used $\hat{r}_{\rm in}=8 m$) and the maximum rest-mass density inside the torus (but not its position).
No assumptions are made regarding the shape or the outer boundary of the torus. 
Solving the equation of hydrostatic equilibrium (\ref{eq:euler}) together with 
(\ref{eq:HC})-(\ref{eq:tfdKijdt}), and (\ref{eq:gauge}) we obtain one black hole-toroid model.
Then we slightly increase the rest-mass density and recompute the same equations. In this way a sequence of
black hole-toroids with increasing mass of the torus is obtained. For low mass ratios and low black hole spins
one model needs $\sim 100$ iterations, while for high mass ratios and high spins $\sim 1500$ iterations are
required. A model is computed when all gravitational and fluid variables have a difference $\sim 10^{-7}$ 
between two successive iterations.

\begin{figure}
\begin{center}
\includegraphics[width=1.05\columnwidth]{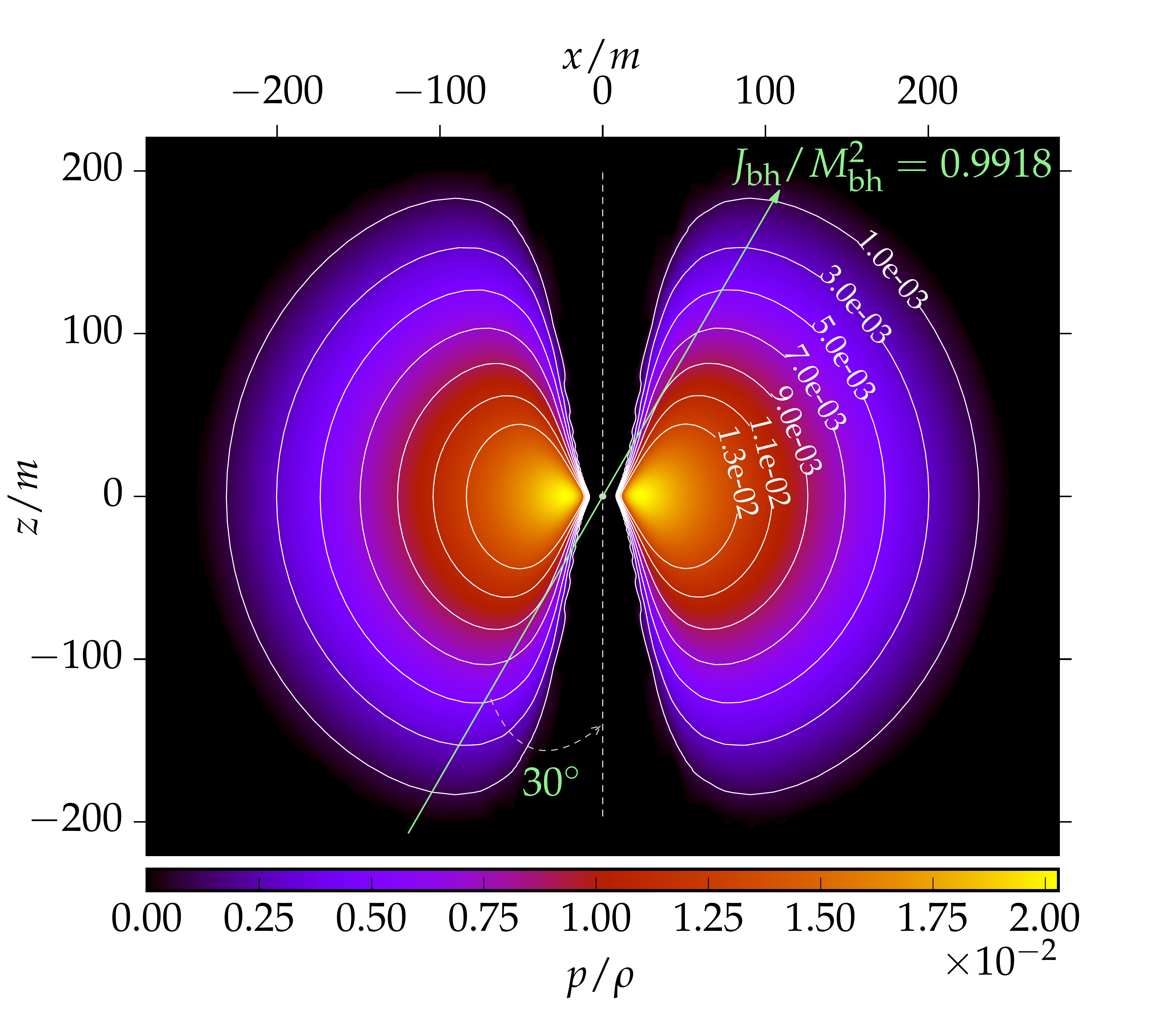}
\caption{Density plot of the function $p/\GR$ (pressure over rest-mass density) on the $x-z$ plane 
for the solution presented in the (\textit{Tilted black-hole-torus system}) section. All units are in 
$G=c=M_\odot=1$. The inner part of the torus corresponds to $\hat{r}_{\rm in}=8 m$.}  
\label{fig:emdg}
\end{center}
\end{figure}

For the particular example shown in Fig. \ref{fig:emdg} we have
black hole parameters, $a/m=0.95$ tilted at an angle $\hat{\GU}=30^{\circ}, \hat{\GP}=0$
(these parameters determine the $3+1$ quantities of the initial background
solution $\GA_{\ks},\ \GB^i_{\ks},\ \GC_{\ks},\ h_{ij}^{\ks}$),
a barotropic EoS with $K=123.6$, $\Gamma=2$, and a rotation law of the form 
$j(\Omega)=A^2 \Omega\left[\left(\frac{\Omega_c}{\Omega}\right)^q-1\right]$
with $A=0.1$, $q=1$ \cite{Uryu16}. 
The two constants that appear in our equations, ($\mathcal{E},\Omega_c$), are computed by evaluating
Eq. (\ref{eq:euler}) at two points (the inner fixed point of the torus and the point of maximum density)
and solving the resulting nonlinear system at each iteration.

The disk has maximum height of $200 m$ and maximum width of $242 m$
with inner point at $\hat{r}=8 m$ as seen in Fig. \ref{fig:emdg}.
Its rest mass is $M_0=5.181 m$.
The apparent horizon of the vacuum Kerr black hole is at $r_{+}/m=1+\sqrt{1-(a/m)^2}=1.312$.
It intersects the x-axis
at $\hat{r}_{+}=1.5233 m$ while its intersection with the 
z-axis happens at $\hat{r}_{+}=1.3726 m$. Our grid covers the region $\hat{r}\in[1.2498, 10^5] m$
and all angles, and excises the region $\hat{r}<1.2498$.
%The rest mass of the torus is  $M_0=5.181 m$ 
%and its angular momentum $J_{\rm t}^i=(0,0,30.17) m^2$.
The system has Arnowitt-Deser-Misner (ADM) mass $M=6.144 m$ and angular momentum
$J^i=(0.4908, -0.001519, 31.19) m^2$.
%The angular momentum of the black hole is 
%$J^i_{\rm bh} = J^i-J_{\rm t}^i=(0.4908, -0.001519, 1.019) m^2$
%with magnitude $J_{\rm bh}=1.131 m^2$.

The angular momentum of the black hole is calculated through the isolated horizon
formalism \cite{Dreyer03}, $J^i_{\rm bh} =(0.5169, -0.0006792, 0.8925) m^2$.
Using the apparent horizon finder described in \cite{Tsokaros07} we calculated the mass
of the black hole \cite{Christ70} to be $M_{\rm bh}=1.0198m$.
Thus the dimensionless spin of the black hole in the black hole-toroid system is 
$J_{\rm bh}/M_{\rm bh}^2=0.9918$. The angular momentum of the torus is then
$J^i_{\rm t} = J^i-J_{\rm bh}^i=(-0.02614, -0.0008393, 30.29) m^2$. If one uses a
Komar integral to calculate the angular momentum of the torus the result is
$J_{\rm t,Komar}^i=(0,0,30.17) m^2$ which shows good agreement in the z-component.
%and  therefore its dimensionless
%spin $J_{\rm bh}/M_{\rm bh}^2=0.9993$. 
This model was the last member of a sequence of black hole-toroids
with increasing rest-masses starting from %$M_0\sim 10^{-4}m$ and $J_{\rm bh}/M_{\rm bh}^2=0.95$. 
an infinitesimal disk of rest mass $\sim 10^{-4}m$ around a Kerr black hole of dimensionless spin 
$a/m=0.95$.
As the torus gains mass and angular momentum it spins-up the black hole. The last model computed here
with $M_0=5.181 m$ has spinned up the black hole to almost maximal spin. A further increase in the
angular momentum and mass of the torus in a quasiequilibrium state is impossible since it will 
drive the spin of the black hole beyond the maximum value.

%%%%%%%%%%%%%%%%%%%%
%%%  Discussion  %%%
%%%%%%%%%%%%%%%%%%%%
\section{Discussion}

In this work we present a new formulation for the initial value problem in general relativity for spacetimes
that contain a black hole and the first nonaxisymmetric black hole-disk solution. Here the disk is $\sim 5$ 
times more massive than the black hole and the hole has near-extremal spin.

Our formulation provides a good starting point for numerical evolution calculations. Unlike other methods
it does not assume a conformal metric (6 components) but instead 3 gauge conditions (3 components) chosen to
match known, closely related, physical models (e.g. Kerr-Schild black holes or axisymmetric stars). For stationary
axisymmetric spacetimes our formulation yields the unique equilibrium solutions. 
%For nonaxisymmetric spacetimes our solutions are not equilibria, but \textit{they generate no violations} when 
%substituted into the full Einstein system (e.g. $[\pd_t K_{ij}]^{\tf}$) by construction, in contrast to other 
%commonly adopted formulations.
For nonaxisymmetric spacetimes our solutions are not equilibria, but in contrast to other commonly adopted
formulations, they provide a way of controlling the gravitational wave content in a self-consistent way.

Although in the present article we used excision, it would not be difficult using the same decompositions to 
solve also for puncture initial data (by decomposing the conformal factor and solving the Hamiltonian for the 
regular part), which are widely also used.
We think that our method will be useful in the gravitational wave detection-multimessenger astronomy
era since it can compute more accurate initial values needed for simulations similarly to what the original 
waveless formulation did for binary neutron stars \cite{Uryu06,Uryu09}.
Problems such as junk radiation, better imposition of helical symmetry,
or more accurate resolution of tidal effects are
examples where our new method can be more appropriate than current studies.

%%%%%%%%%%%%%%%%%%%%%%                                                           
%   Acknowledgments                                                              
%%%%%%%%%%%%%%%%%%%%%%                                                                               
\acknowledgements 
We thank M. Ruiz for providing the massless disk model.
This work was supported by NSF Grants No. PHY-1602536 and No. PHY-1662211 and NASA grant
80NSSC17K0070 to the University of Illinois at Urbana-Champaign, as well as by 
JSPS Grant-in-Aid for Scientific Research (C) 15K05085 and 18K03624 to the University of Ryukyus. 
A.T. would like to thank the University of Ryukyus for their hospitality during a visit.

%%%%%%%%%%%%%%%%%%%%%%%%%%%%%%%%%%%%%%%%
\bibliographystyle{apsrev4-1}        %%%
%\bibliography{references}            %%%

%%%%%%%%%%%%%%%%%%%%%%%%%%%%%%%%%%%%%%%%

\end{document}